\documentclass{article}
\usepackage{graphicx} 
\usepackage[hyphens,spaces,obeyspaces]{url} 
\usepackage{authblk} 
\usepackage{orcidlink} 
\usepackage{booktabs}
\usepackage{tabularx} 
\usepackage{float}
\usepackage{csquotes} 
\usepackage[natbibapa]{apacite} 
\usepackage{multirow}

\title{Presenting a classifier to detect research contributions in OpenAlex}
\author{Nick Haupka \orcidlink{0009-0002-6478-6789}}

\affil{Göttingen State and University Library, University of Göttingen}
\affil{Platz der Göttinger Sieben 1, 37073 Göttingen, Germany}
\affil{\href{mailto:nick.haupka@sub.uni-goettingen.de}{nick.haupka@sub.uni-goettingen.de}}

\date{}

\begin{document}

\maketitle

\section*{Abstract}
This paper introduces a document type classifier with the purpose to optimise the distinction between research and non-research journal publications in OpenAlex. Based on open metadata, the classifier can detect non-research or editorial content within a set of classified articles and reviews (e.g. paratexts, abstracts, editorials, letters). The classifier achieves an F1-score of 0,95, indicating a potential improvement in the data quality of bibliometric research in OpenAlex when applying the classifier on real data. In total, 4.589.967 out of 42.701.863 articles and reviews could be reclassified as non-research contributions by the classifier, representing a share of 10,75\%. 

\bigskip
\noindent\textbf{Keywords}: OpenAlex, Document types, machine learning, classification, bibliometrics

\section{Introduction}
Bibliometric databases like OpenAlex \citep{priem_openalex_2022}, Crossref \citep{hendricks_crossref_2020} and Semantic Scholar \citep{kinney_semantic_2023} index millions of research publications each year. In this process, the document type plays a decisive role in distinguishing different types of works such as editorials, case reports, journal articles and reviews. The decisions made by database providers to classify publications into document types have a profound impact on numerous bibliometric procedures and performance evaluations. For example, university rankings like the Leiden Ranking Open Edition \citep{van_eck_methodology_2024} restrict their performance analysis on certain document types such as articles and reviews, while excluding document types like editorials and letters. Another measurement that relies on document types given by scholarly databases is the Journal Impact Factor. The Journal Impact Factor represents the influence of a journal by counting citations of articles and reviews published in a journal\footnote{\url{https://support.clarivate.com/ScientificandAcademicResearch/s/article/Journal-Citation-Reports-Document-Types-Included-in-the-Impact-Factor-Calculation?language=en_US}}. Document types are also frequently used as a restriction criterion for research publications, for example when analysing the dissemination of open access in scholarly communication \citep{piwowar_state_2018} or when examining the influence of transformative agreements on hybrid journals \citep{jahn_how_2025}. 

However, the typology to describe works in bibliometric databases as well as the methods used to assign document types vary between data sources and providers. A study by \cite{haupka_analysis_2024} shows that OpenAlex tends to overestimate the assignment of the document type \textit{article} when comparing it to Scopus, Web of Science, Semantic Scholar and PubMed. \cite{alperin_analysis_2024} made a similar observation, noting deviations in the classification of document types between OpenAlex and Scopus. Studies also show that the methods for applying document type classification by scholarly data providers are not always precise as the document type assigned within a database can sometimes differ from the actual document type. For instance, \cite{donner_document_2017} investigated a sample of document types retrieved from Scopus and Web of Science and compared them to the official publisher websites and found discrepancies in more than 17\% of the cases. Sometimes, document types are also not covered in bibliographic databases. For example, an analysis of data journals by \cite{jiao_how_2023} has shown that OpenAlex lacks the classification of data papers which are assigned to an independent document type in Scopus and Web of Science but considered as journal articles in OpenAlex. \cite{delgado-quiros_completeness_2024} compared typologies of document types of eight different bibliometric databases, including the databases OpenAlex, Crossref, The Lens, Scilit and Dimensions and found a high degree of consistency between document types retrieved from these databases, which can be attributed to Crossref as the main source of all databases mentioned. 

In the past year, the team behind OpenAlex, OurResearch, has announced and implemented several changes to its document type classification\footnote{\url{https://github.com/ourresearch/openalex-guts/blob/main/files-for-datadumps/standard-format/RELEASE_NOTES.txt}}. First and foremost, it included the classification of preprints and reviews. Further it integrated document types from PubMed to relabel a set of articles as editorials and letters. Also, it extended their list-based approach for detecting paratexts and editorial material. The effects of this changes were documented by \cite{haupka_recent_2024}. While the implementation of document types from PubMed reduces the amount of works with the type \textit{article}, a gap between the document type classification of OpenAlex and proprietary databases like Scopus and Web of Science still remains.

In recent years, the application of machine learning algorithms have become increasingly popular in open science and especially in bibliometric databases \citep{jeangirard_lutilisation_2022}. A prominent example is Semantic Scholar which uses machine learning classifiers for, among others, author disambiguation and field of study classification \citep{kinney_semantic_2023}. Another example is OpenAlex which uses a probability model to detect the language of a work. Further it uses machine learning classifier to match records with topics and SDGs\footnote{\url{https://docs.openalex.org/api-entities/works/work-object}}. Also, the French Open Science Monitor uses machine learning algorithm to detect datasets and software \citep{bracco_how_2025}.

In line with these developments, I have developed a document type classifier that filters research contributions in OpenAlex based on open metadata such as author count and number of references \citep{haupka_identifying_2024}. Similar efforts were made by \cite{charalampous_classifying_2017} who created a classifier that takes metadata into account to distinguish between papers, theses and slides in (open access) repositories. Metadata for the classifier were extracted and aggregated from PDFs. They consisted of the number of authors, the total number of words in a document, the number of pages and the average number of words per page. Overall, this approach resulted in a high F1-score (0.96). A similar approach was conducted by \cite{caragea_document_2016} who extracted file and text specific features from documents to differentiate between books, slides, theses and papers using machine learning techniques. A more traditional approach to address the issue of misclassification of journal articles was proposed by \cite{maisano_large-scale_2025}. It uses a concordance matrix to identify misclassified documents between two bibliographic databases (in this case Web of Science and Scopus). The methodology involves a manual control of document types which makes it difficult to apply on large amounts of journal publications. 

In this paper, I want to describe my classifier and my approach in depth. My results were previously briefly documented in a blog post in October 2024\footnote{\url{https://subugoe.github.io/scholcomm_analytics/posts/oal_document_types_classifier/}}. Since then, I implemented minor changes to the classifier. Therefore, I wanted to update my results.

\section{Data}
Creating a gold standard dataset for validating document types is difficult. Data sources like Crossref, OpenAlex and PubMed mostly rely on publisher information when classifying document types\footnote{\url{https://crossref.gitlab.io/knowledge_base/docs/topics/content-types/}}\footnote{\url{https://pubmed.ncbi.nlm.nih.gov/help/}}. As \cite{haupka_analysis_2024} has demonstrated, this can lead to errors in databases because each publisher can decide on his own how he classifies a publication. Also, data sources like Crossref and PubMed apply different taxonomies which makes it difficult to compare document types from different data sources. 

For this analysis, I decided to use document types from PubMed as my previous analysis has shown that the data source reaches comparable accuracy against Scopus and Web of Science in regard to the classification of research articles and reviews \citep{haupka_analysis_2024}. Because, I wanted to validate my classifier results on other data sources like Semantic Scholar or Scopus in a later phase, I constructed a shared corpus between OpenAlex, Scopus, Web of Science, Semantic Scholar and PubMed with limitation to the publication type journal and the publication years 2012 to 2022. 

Document types from PubMed were reassigned to one of two classes: research and non-research. A complete mapping table can be found in the code supplement\footnote{\url{https://github.com/naustica/openalex_doctypes/blob/classifier/queries/oal_doc_dataset_extended.sql}}. Note that case reports were originally categorised as research. However, after validating the classification results, I came to the conclusion that this type should be categorised as a non-research, as it is too short for a research article or review. 

Analysis data was partially retrieved from the German Competence Network for Bibliometrics \citep{schmidt_data_2024}. This includes data from Scopus (July 2024 snapshot) and OpenAlex (August 2024 snapshot). Besides I used data from OpenAlex, Crossref and PubMed to train the classifier. The results of the classifier, including scores, are made available on Zenodo \citep{haupka_2025_15308680}. 

As of now, OpenAlex includes 17 types of documents, when restricting to the primary location source type \textit{journal} and publication years 2014 to 2023. Research articles and reviews are the most prominent types with overall 93,9\%. In all cases, a document type was linked to a work. Note that 8.445.541 (11,27\%) of works do not have a source type in OpenAlex (for the publication years 2014 to 2023). 

\section{Approach}
As \cite{charalampous_classifying_2017} I made the assumption that document types have specific metadata characteristics which can be used to differentiate distinct types of works. Here I also rely on the analysis of \cite{haupka_analysis_2024} who demonstrated that works that are categorised as research contributions (e.g. reviews and articles) have, among other metadata, different average numbers of authors and references as non-research contributions (e.g. editorials, paratexts and letters). 

My experiments resulted in ten features (F1-F10) that I have chosen to train the classifier (see Table \ref{tab:features}). Some of these features can be taken from the databases (Crossref and OpenAlex) without further problems while others have to be manually calculated. This applies to F2, F3 and F4.

\begin{table}[ht]
\centering
\makebox[\textwidth]{
\begin{tabularx}{1.3\textwidth}{lXXX}
\toprule
\textbf{Feature} & \textbf{Type} & \textbf{Retrieved from} & \textbf{Manually \newline calculated} \\
\hline
F1: has abstract? & Boolean & Crossref & False \\
F2: title word count & Integer & Crossref & True \\
F3: page count & Integer & Crossref & True \\
F4: author count & Integer & Crossref & True \\
F5: has license? & Boolean & Crossref & False \\
F6: number of citations & Integer & Crossref & False \\
F7: number of references & Integer & Crossref & False \\
F8: has funding information? & Boolean & Crossref & False \\
F9: number of affiliations & Integer & OpenAlex & False \\
F10: has OA url? & Boolean & OpenAlex & False \\
\bottomrule
\end{tabularx}}
\caption{Selected features for the classifier.}
\label{tab:features}
\end{table}

I decided to integrate and train the classifier on some bibliographic data from Crossref instead of OpenAlex, as Crossref covers certain metadata fields more comprehensively, especially when looking at the provision of page numbers \citep{delgado-quiros_completeness_2024}. I calculated the page number by computing the absolute number of the difference of the first page and the last page. In cases where the first or last page number was missing I decided to give the document a page count of 1. Note that I have applied regular expressions on the page strings to obtain better results when calculating the page numbers. An additional filter was applied in section 6, namely: If the journal issue contains the words \enquote{\%sup\%} (for supplementary materials) or \enquote{\%meet\%} (for meeting abstracts) the publication was considered as non-research. This affects about 803.688 (1,88\%) articles and reviews in the dataset.

\section{Experiments}
Experiments were carried out using the following machine learning algorithms: Logarithmic Regression (LR), Random Forest (RF), K-nearest-neighbours (KNN) and AdaBoost with Decision Tree (AdaBoost). In addition, I used random class assignment as a baseline (Baseline). Data were split into 80\% training data, 10\% test data and 10\% validation data. Since research publications are more prominent in the dataset than non-research publications with over 92,58\%, I have stratified the class distribution when splitting the data into training, test and validation set. In order to reduce the size of the dataset, publications from publishers with fewer than 5000 publications were removed. I used grid search for optimising hyper-parameters of each individual model. The results of my experiments can be seen in Table \ref{tab:results}.

\section{Results}

Overall, the selected models perform very well on the dataset as can be seen in Table \ref{tab:results}. The random forest classifier achieves the highest precision on both, the test dataset and the validation set, when comparing to the other algorithms with 94.8\%. The k-nearest-neighbours algorithm achieves the highest recall and also the highest F1-score (both close to 95\%), also on both, the test and validation set. 

\begin{table}[ht]
\centering
\makebox[\textwidth]{
\begin{tabularx}{1.2\textwidth}{clXXXXX}
\toprule
\multicolumn{7}{c}{Algorithm} \\
\textbf{} & Measure & LR & RF & KNN & AdaBoost & Baseline \\
\hline
\multirow{3}{*}{\parbox{1.2cm}{Test Results}}
& Precision & 0.9429 & \textbf{0.9482} & 0.9470 & 0.9269 & 0.8631 \\
& Recall & 0.8504 & 0.9433 & \textbf{0.9481} & 0.9352 &  0.5000 \\
& F1-score & 0.8804 & 0.9454 & \textbf{0.9475} & 0.9297 & 0.6108 \\
\hline
\multirow{3}{*}{\parbox{1.5cm}{Validation Results}}
& Precision & 0.9427 & \textbf{0.9480} & 0.9466 & 0.9263 & 0.8627 \\
& Recall & 0.8499 & 0.9429 & \textbf{0.9477} & 0.9346 & 0.4997 \\
& F1-score & 0.8801 & 0.9451 & \textbf{0.9471} & 0.9292 & 0.6105 \\
\bottomrule
\end{tabularx}}
\caption{Test and validation set results}
\label{tab:results}
\end{table}

Table \ref{tab:results_classes} shows the results of each model for the chosen classes research and non-research. As can be seen, the results are quite diverse as each model shows advantages and disadvantages. For example, Logarithmic Regression outperforms all other models when classifying research contributions with over 99\% precision. Further, it achieves the highest recall among the other models when predicting non-research publications. However, it performs poorly when detecting non-research publications (only about 32\% precision). Also, in comparison with the other models, it achieves a lower recall with only about 84\% when predicting the research class. The highest precision for predicting non-research publication is achieved by the k-nearest-neighbours algorithm (about 65\%). Along with the highest F1-score for predicting the research class, the k-nearest neighbours algorithm was chosen for productive use.

\begin{table}[ht]
\centering
\makebox[\textwidth]{
\begin{tabularx}{1.2\textwidth}{clXXXXX}
\toprule
\multicolumn{7}{c}{Algorithm} \\
\textbf{} & Measure & LR & RF & KNN & AdaBoost & Baseline \\
\hline
\multirow{3}{*}{\parbox{1.2cm}{Research}}
& Precision & \textbf{0.9924} & 0.9761 & 0.9699 & 0.9536 & 0.9258 \\
& Recall & 0.8443 & 0.9618 & 0.9737 & \textbf{0.9769} &  0.4997 \\
& F1-score & 0.9124 & 0.9689 & \textbf{0.9718} & 0.9651 & 0.6491 \\
\hline
\multirow{3}{*}{\parbox{1.2cm}{Non-research}}
& Precision & 0.3209 & 0.5967 & \textbf{0.6546} & 0.5847 & 0.0740 \\
& Recall & \textbf{0.9193} & 0.7062 & 0.6228 & 0.4065 & 0.4998 \\
& F1-score & 0.4757 & \textbf{0.6469} & 0.6383 & 0.4796 & 0.1289 \\
\bottomrule
\end{tabularx}}
\caption{Classifier results for each class based on test set}
\label{tab:results_classes}
\end{table}

\section{Validation}
In this section, I will apply the classifier to real data and compare it to other data sources. Firstly, I will compare with OpenAlex to see the impact of the classification results. Secondly, I will compare the classifier with Scopus. Finally, I will test two small samples of one hundred publications each to evaluate the accuracy of the classifier. 

\subsection{Comparison with OpenAlex}
Overall, the classifier recognised 4.589.967 out of 42.701.863 articles and reviews from journals between 2014 and 2023 as non-research publications. This corresponds to about 10,75\%. When applying the classification results to OpenAlex topics, one can see that publications from the Health and Social Sciences were more frequently classified as non-research contributions than publications from Physical and Life Sciences. This is probably due to many case reports and abstracts in the Health Sciences which usually do not contain references and are often one or two pages long. This observation is also made when assigning the classification results to institutional types from the Research Organization Registry (ROR)\footnote{\url{https://ror.org}}, which are included in OpenAlex. Here, the institutional types healthcare and company are more frequent than the types education and government for non-research publications.

\begin{figure}[H]
\centering
\makebox[\textwidth]
    {\includegraphics[width=14cm]{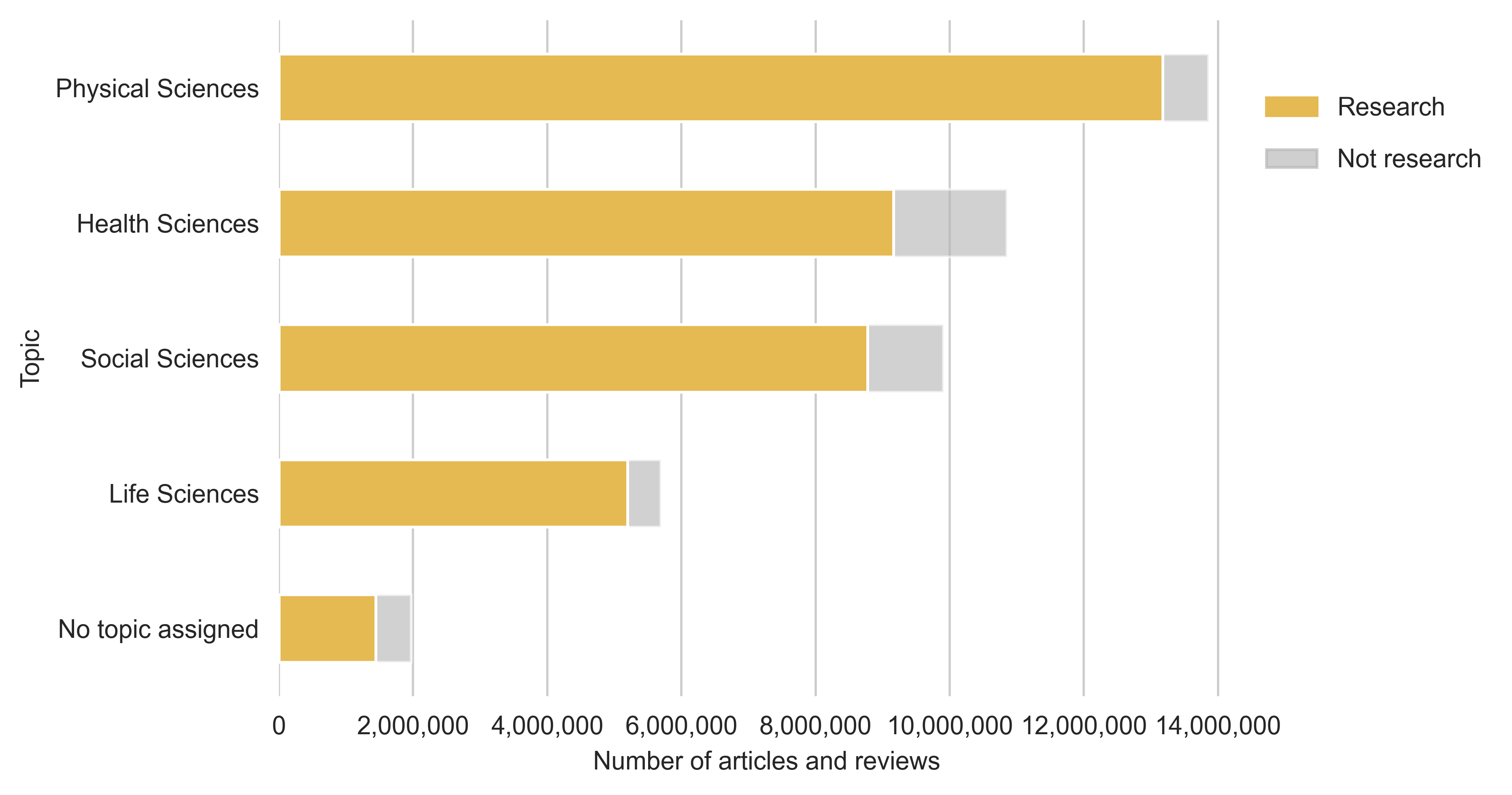}}
\caption{\label{fig:topics}Classifier results applied on topics in OpenAlex.}
\end{figure}

\subsection{Comparison with Scopus}
When evaluating the classifier on a shared corpus based on DOI matching between OpenAlex and Scopus and using solely the document type classification of Scopus, it shows that only 327.522 out of 22.338.918 articles and reviews in journals between 2014 and 2023 are labeled as non-research publications by the classifier. This corresponds to about 1,5\%.

\begin{figure}[H]
\centering
\makebox[\textwidth]
    {\includegraphics[width=14cm]{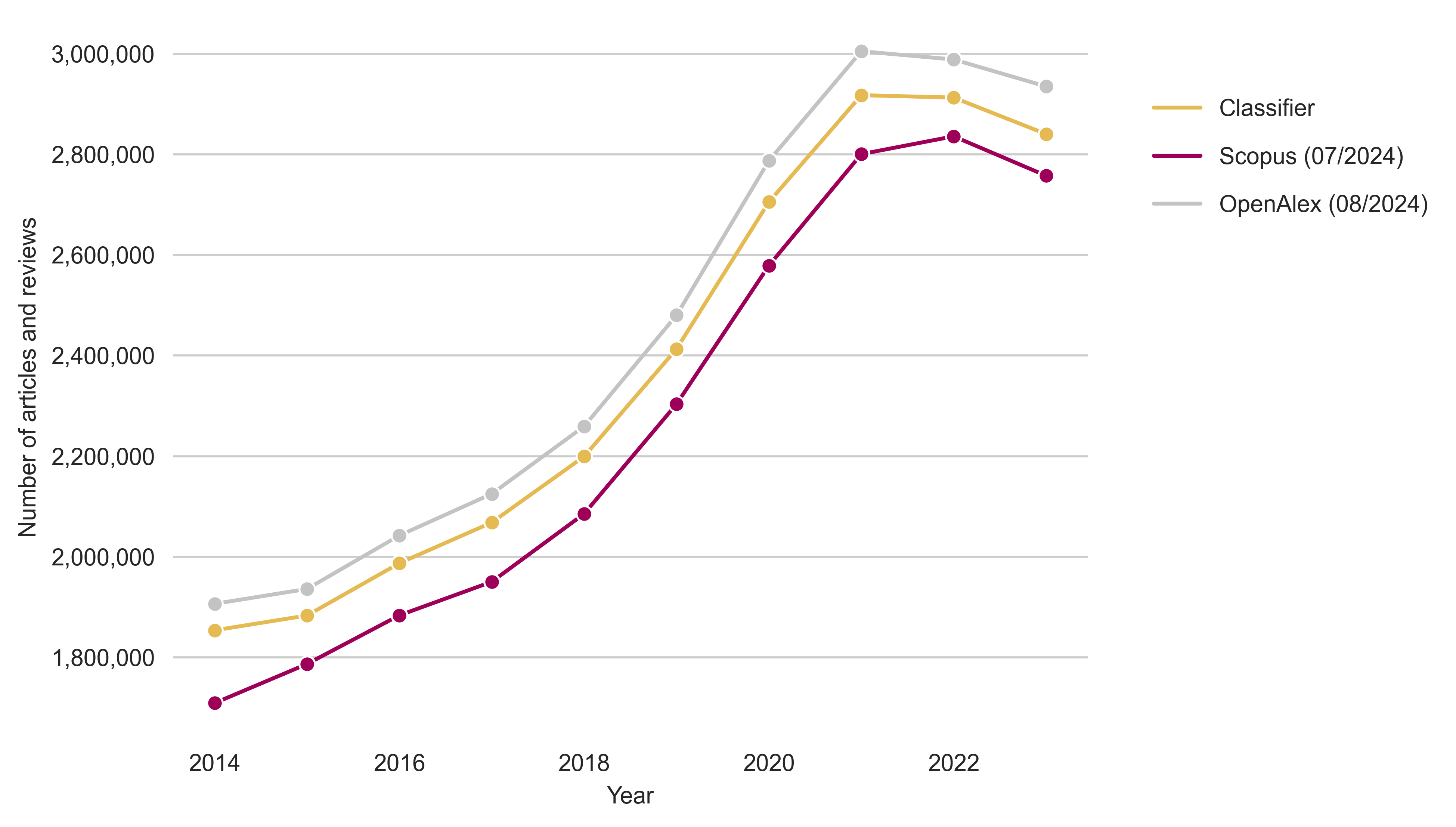}}
\caption{\label{fig:oal_scp_comparison}Comparison of the classification of articles and reviews in journals between OpenAlex, Scopus and the introduced classifier based on a shared corpus.}
\end{figure}

Figure \ref{fig:oal_scp_comparison} shows a comparison of the coverage of articles and reviews in journals restricted to the publication years 2014 to 2023 between OpenAlex, Scopus and the introduced classifier. While OpenAlex classifies more than 24.465.047 (95,87\%) of publications in the shared corpus as articles and reviews, Scopus only counts 22.691.562 (88,92\%) items as research. The classifier is positioned between the two and counts 23.781.060 (93,19\%) items as research.

\subsection{Manual check}
I created two cases of samples that I want to check manually:

\begin{itemize}
\item Sample 1 (n=100): Articles and reviews in journals that are identified as research contribution by the classifier.
\item Sample 2 (n=100): Articles and reviews in journals that are identified as non-research contribution by the classifier.
\end{itemize}

Overall, 75\% of items were correctly classified as research in Sample 1. 16\% of items could not be checked due to not resolving DOIs or not loading websites. Also some items required a login. 9\% of items were falsely classified as research. Some of them were conference articles, abstracts, corrections or book reviews. 
For Sample 2, I only considered articles and reviews that were classified by the classifier as non-research and were actually not an article or review (e.g. editorial, abstract or paratext) as correct whereas all article and reviews that were falsely classified by the model were considered as incorrect. Overall 69\% of the results were classified correctly as non-research. In contrast, 21\% were falsely classified as non-research. However, several of them originated from Malaysian, Indoniesian or Russian sources which could indicate a lack of provision of metadata by certain foreign (non-English) publishers or websites. Further 10\% could not be checked, because resources were not found on the corresponding website or the corresponding website did not respond. 

\subsection{Examples of classified data}
\begin{figure}[H]
\centering
\makebox[\textwidth]
    {\includegraphics[width=8cm]{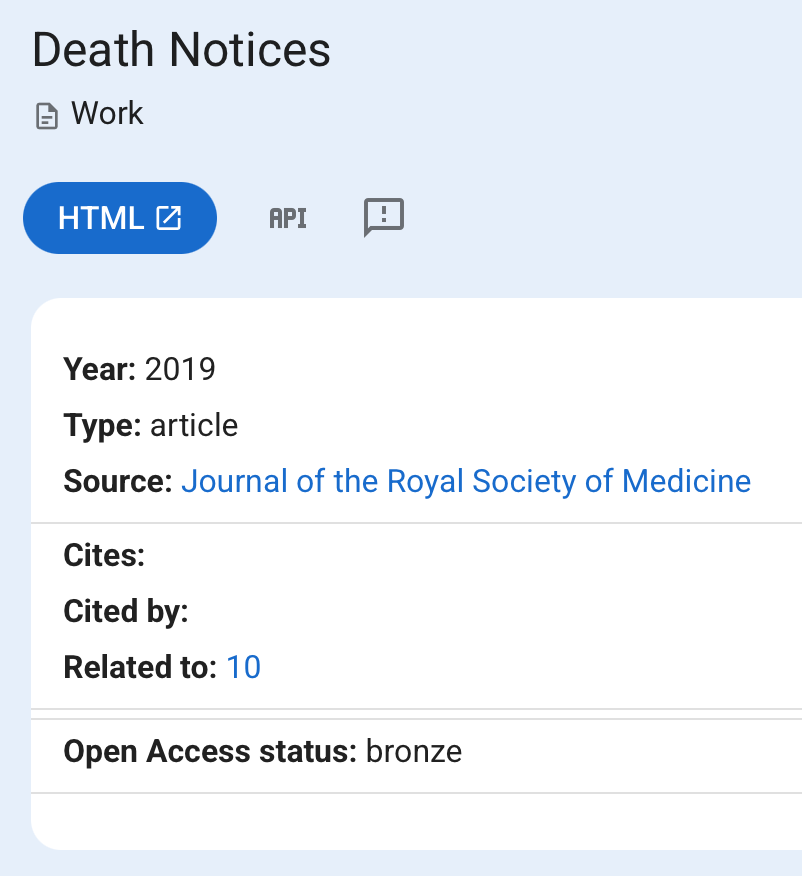}}
\caption{\label{fig:paratext1}Article that was classified as non-research by the classifier (\url{https://openalex.org/W4206081200})}
\end{figure}

\begin{figure}[H]
\centering
\makebox[\textwidth]
    {\includegraphics[width=10cm]{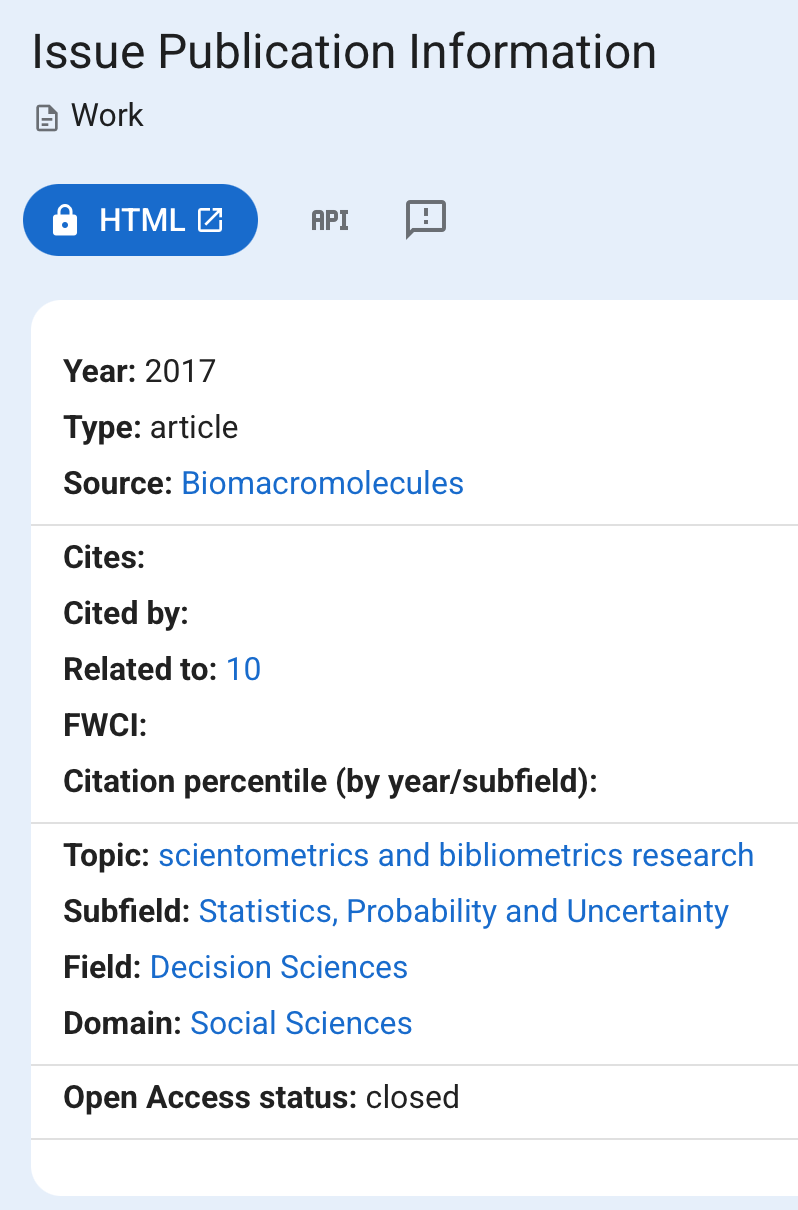}}
\caption{\label{fig:paratext2}Article that was classified as non-research by the classifier (\url{https://openalex.org/W4237974798})}
\end{figure}

\begin{figure}[H]
\centering
\makebox[\textwidth]
    {\includegraphics[width=16cm]{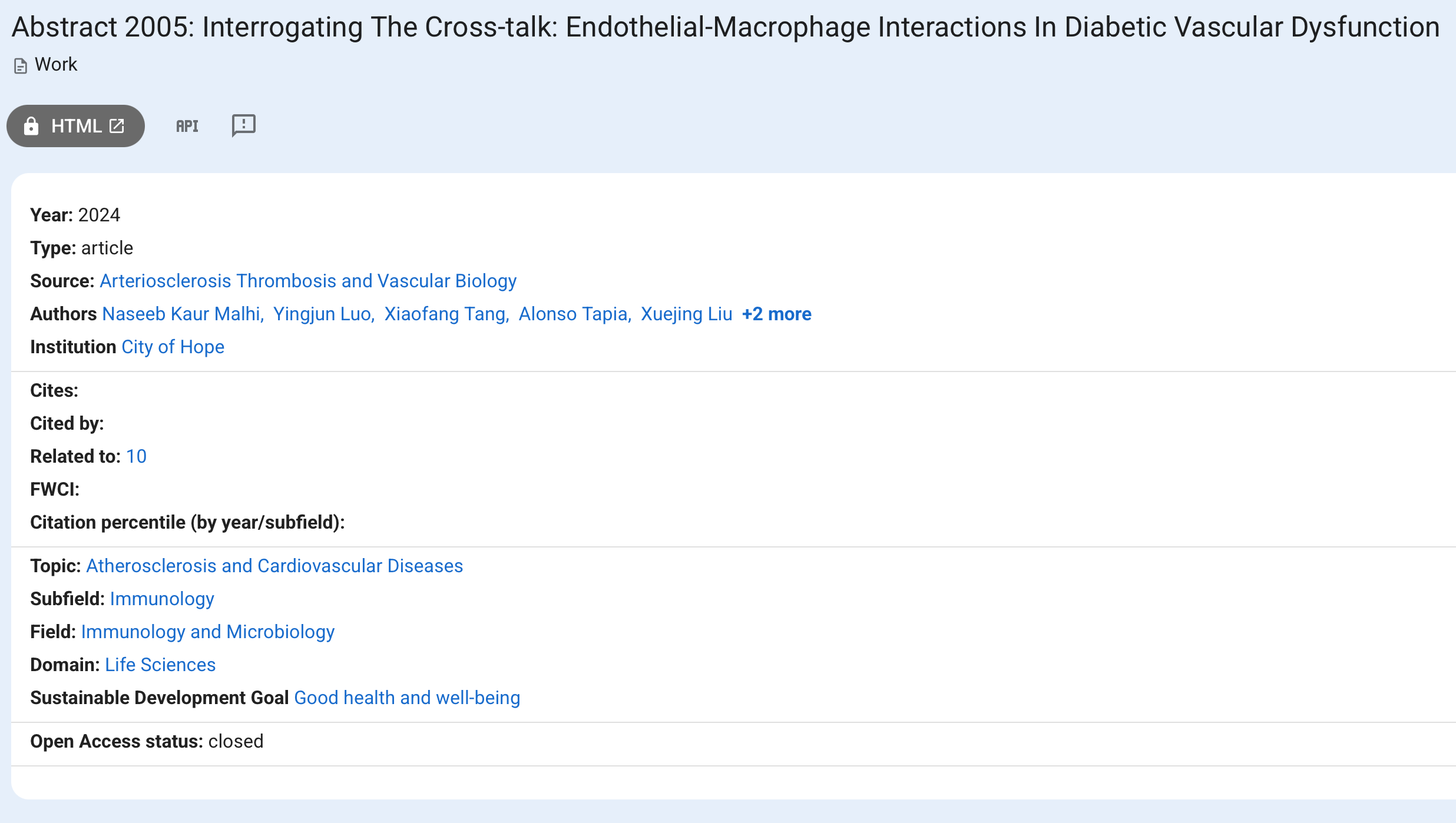}}
\caption{\label{fig:paratext3}Article that was classified as non-research by the classifier (\url{https://openalex.org/W4400934704})}
\end{figure}

In this section, I provide three examples of publications that were classified as articles by OpenAlex and were detected by the classifier as non-research contributions (see Figure \ref{fig:paratext1}, \ref{fig:paratext2}, \ref{fig:paratext3}). All examples were requested on February 6, 2025. 

\section{Discussion}
My classification approach provides a fast, scalable and simple method to detect and distinguish research from non-research publications in OpenAlex. With the help of the classifier, over 10\% of articles and reviews can be classified as non-research contributions. In contrast, the classifier only detects around 1,5\% of articles and reviews in Scopus as non-research based on a shared corpus between Scopus and OpenAlex. The sample analysis shows that the classifier reaches around 89\% accuracy for detecting research contributions and around 75\% accuracy on classifying non-research works (when excluding resources that were not available, e.g. not resolving DOIs). This observation is in line with metrics that were achieved in the training process of the model. I would also like to emphasise that the model performs well across works of different languages. I find an even distribution when it comes to the classification of document types from English and non-English publications although English publications are predominantly present in OpenAlex \citep{cespedes_evaluating_2025}. In contrast, a language-based approach that relies on the title and abstract of a publication has the disadvantage that abstracts are often not available to the required extent and that a title cannot always be meaningful (if the title suggests a research article but is ultimately only an abstract). A full-text-based approach, in which content is extracted from PDFs, could be more helpful here (e.g. \cite{caragea_document_2016, charalampous_classifying_2017}). When comparing my approach with other existing methods, for instance the Leiden Ranking Open Edition, it shows that the classifier excludes fewer publications which can be attributed to the exclusion of journals and publications with missing metadata in the Leiden Ranking \citep{van_eck_methodology_2024, haupka_identifying_2024}. In contrast to semi-automated processes, where manual checks are necessary (e.g. \cite{maisano_large-scale_2025}), the classifier can also be applied to large amounts of data, making it more suitable for extensive analyses. However, the classifier can also be combined with manual validation. A noteworthy example of a platform for manual validation of metadata in OpenAlex is the Works-magnet developed by the French Ministry of Higher Education and Research\footnote{\url{https://works-magnet.esr.gouv.fr}}. It encourages users to correct automatically computed results in OpenAlex which than will be applied in the database. \cite{mokhnacheva_document_2023} also suggests manual validation of document types using original information from publishers.

But my approach also comes with some limitations:

\begin{itemize}
\item Coverage of metadata: I noticed significant discrepancies between metadata of different publishers and sources. While some publishers provide helpful metadata such as references and abstracts, others do not, which results in a lower accuracy of the classifier. Further, problems of missing metadata exist, e.g. in the cases of institutions \citep{zhang_missing_2024}, page numbers \citep{delgado-quiros_completeness_2024} and funding information \citep{culbert_reference_2025}.
\item Quality of metadata: I noticed errors for some publications that have a wrong citation count. For example, I found several paratexts that have a high citation count although under investigation it shows that the citations are not correct\footnote{\url{https://openalex.org/works?page=1&filter=type\%3Atypes\%2Farticle,display_name.search\%3Atitelei}}. Some page numbers are also not calculated correctly, either because OpenAlex contains incorrect page numbers or the page numbers are not available in a suitable format for calculation (e.g. containing letters). 
\item Quality of training set: Although PubMed has a quality comparable to Scopus and Web of Science in the classification of document types, I cannot guarantee that the training set does not contain errors. 
\end{itemize}

Although the classifier was originally developed to recognise editorial content within a classified set of journal articles and reviews, I have found that many of the classification results indicate that OpenAlex also classifies a large number of abstracts, case reports and book reviews as journal articles. Here, the classifier can be used to identify journals that have a disproportionate share of non-research publications. This can help to check whether these journals have a focus on research articles or not. For example, the journal Reactions Weekly predominantly publishes case reports which however are labelled as articles in OpenAlex\footnote{\url{https://link.springer.com/journal/40278}}. Another journal is the Bulletin of the Center for Children's Books, which mostly publishes book reviews that are classified as articles in OpenAlex\footnote{\url{https://bccb.ischool.illinois.edu}}

For the future, I would like to explore more rule-based decisions that can be applied on top of the classifier. For example, by extending the title heuristics of OpenAlex with additional phrases like \enquote{Table of Contents} or \enquote{Meeting Abstract}. I would also like to explore how certain hard rules could affect the analysis of OpenAlex data, for example excluding all publications with an author count of 0. 

\newpage

\section*{Competing interests}
The author declares no conflict of interest.

\section*{Funding Information}
This work was funded by the Federal Ministry of Education and Research (grant funding number: 16WIK2301E, The OpenBib project).

\section*{Code Availability}
The Jupyter notebook containing the source code analysis can be found on GitHub: \url{https://github.com/naustica/openalex doctypes}.

\section*{Author Contributions}
Conceptualization, investigation and Writing - original draft : Nick Haupka

\newpage
\bibliographystyle{apacite}
\bibliography{references}

\end{document}